\documentclass{amsart}
\usepackage{amssymb}

\usepackage{graphicx}
\usepackage{amscd}

\newtheorem{theorem}{Theorem}
\theoremstyle{plain}
\newtheorem{acknowledgement}{Acknowledgement}

\newtheorem{claim}{Claim}

\newtheorem{corollary}{Corollary}

\newtheorem{lemma}{Lemma}

\numberwithin{equation}{section}

\input{tcilatex}

\begin{document}
\title[Generalizations of Kochen and Specker's Theorem]{Generalizations of Kochen and Specker's Theorem and the Effectiveness of
Gleason's Theorem}
\author{Ehud Hrushovski}
\address{Department of Mathematics, the Hebrew University}
\email{ehud@math.huji.ac.il}
\author{Itamar Pitowsky}
\address{Department of Philosophy, The hebrew University}
\email{itamarp@vms.huji.ac.il}
\urladdr{http://edelstein.huji.ac.il/staff/pitowsky/}

\begin{abstract}
Kochen and Specker's theorem can be seen as a consequence of Gleason's
theorem and logical compactness. Similar compactness arguments lead to
stronger results about finite sets of rays in Hilbert space, which we also
prove by a direct construction. Finally, we demonstrate that Gleason's
theorem itself has a constructive proof, based on a generic, finite,
effectively generated set of rays, on which every quantum state can be
approximated.
\end{abstract}

\maketitle

\section{Gleason's Theorem and Logical Compactness}

Kochen and Specker's (1967)\ theorem (KS)\ puts a severe constraint on
possible hidden-variable interpretations of quantum mechanics. Often it is
considered an improvement on a similar argument derived from Gleason (1957)
theorem (see, for example, Held. 2000). This is true in the sense that KS
provide an explicit construction of a finite set of rays on which no
two-valued homomorphism exists. However, the fact that \emph{there is} such
a finite set follows from Gleason's theorem using a simple logical
compactness argument (Pitowsky 1998, a similar point is made in Bell 1996).
The existence of finite sets of rays with other interesting features also
follow from the same simple consideration. In Pitowsky (1998) some such
consequences, in particular the ``logical indeterminacy principle''are
pointed out, and some are given a direct constructive proof. In this section
we shall formulate the general compactness principle underlying these
results and mention some new ones. In the second section there are some
explicit constructions of finite sets of rays whose existence was inferred
indirectly in the first section; in particular, a simpler proof of the
logical indeterminacy principle. In the last section we prove that there is
an effective (algorithmic) procedure to construct finite sets of rays which
force a uniform approximation to quantum states. In particular, we provide a
short demonstration that Gleason's theorem has a constructive proof, a fact
previously established by Richman and Bridges (1999).

Let $\mathbb{H}$ be a Hilbert Space of a finite dimension $n\geq 3$ over the
complex or real field. A non negative real function $p$ defined on the unit
vectors in $\mathbb{H}$ is called \emph{a state on }$\mathbb{H}$ if the
following conditions hold:

1. $p(\alpha x)=p(x)$ for every scalar $\alpha ,\left| \alpha \right| =1$,
and every unit vector $x\in \mathbb{H}$.

2. If $x_{1},x_{2},...,x_{n}$ is an orthonormal basis in $\mathbb{H}$ then $%
\sum_{j=1}^{n}p(x_{j})=1$.

Gleason's theorem characterizes all states:

\begin{theorem}
\emph{Given a state }$p$\emph{, there is an Hermitian, non negative operator 
}$W$\emph{\ on }$\mathbb{H}$\emph{, whose trace is unity, such that }$%
p(x)=(x,Wx)$\emph{\ for all unit vectors }$x\in \mathbb{H}$\emph{, where }$%
(,)$\emph{\ is the inner product.}
\end{theorem}

Gleason's (1957) original proof of the theorem has three parts: The first is
to show that every state $p$ on $\mathbb{R}^{3}$ is continuous. The second
part is a proof of the theorem in the case of $\mathbb{R}^{3}$, and the
third part is a reduction of the general theorem to $\mathbb{R}^{3}$. The
theorem is also valid in the infinite dimensional case which we shall not
consider.

Let us make more precise what are the formal logical assumptions underlying
the proof of Gleason's theorem. For simplicity we shall concentrate on the
three dimensional real case which comprises the first two parts of Gleason's
proof. All our results are extendable to any real or complex Hilbert space
of a finite dimension $n\geq 3$.

Consider the first order formal theory of the real numbers (that is, a first
order theory of some standard model $\mathbb{R}$ of the reals). This induces
a theory of $\mathbb{R}^{3}$, together with the inner product, and the unit
sphere $\mathbb{S}^{2}$. Add to this first order theory a function symbol $p:%
\mathbb{S}^{2}\rightarrow \mathbb{R}$. Let the Greek letters $\alpha ,\beta
,\gamma ...$ denote variables ranging over the reals and $x,y,z,...$ be
variables ranging over $\mathbb{S}^{2}$. Now add the axioms:

\textbf{G1}. $\forall xp(x)\geq 0$

\textbf{G2}. $\forall x$ $p(-x)=p(x)$.

\textbf{G3}.\emph{\ For each orthonormal triple }$x,y,z\in \mathbb{S}^{2}$%
\emph{\ an axiom: }$p(x)+p(y)+p(z)=1$

Note that in \textbf{G3} we do not use the universal quantifier. Instead 
\textbf{G3} is an axiom schema with a continuum of propositions. The next
axiom is just the statement that every set of reals which is bounded from
below has an infimum. However, since we want to use only first order
formulae we write it as an axiom schema:

\textbf{G4}. \emph{For every one place predicate of reals }$A(.)$\emph{\
expressible in our language an axiom}

$\exists \beta \forall \alpha (A(\alpha )\rightarrow \alpha \geq \beta
)\longrightarrow $ $\exists \beta \lbrack \forall \alpha (A(\alpha
)\rightarrow \alpha \geq \beta )\wedge \forall \varepsilon >0\exists \gamma
(A(\gamma )\wedge \beta >\gamma -\varepsilon )]$

Thus, for example, the claim that $p$ itself has an infimum will follow from 
\textbf{G1} and the application of \textbf{G4} to the predicate $\exists
x(p(x)=\alpha )$. As a matter of fact, the proof of Gleason's theorem
requires twice the application of \textbf{G4}, that is, for two predicates $%
A(.)$.

The proof that $p$ is continuous depends on the axioms \textbf{G1-G4}. One
can see this from Gleason's (1957) original proof , or more directly from
Pitowsky (1998) that only a finite number of application of the schema 
\textbf{G3} are required.

To prove the second part, that every state on $\mathbb{R}^{3}$ is given by a
self adjoint, non-negative, trace one operator, another axiom is needed:

\textbf{G5}.\emph{If }$p$\emph{\ is continuous then its minimum and maximum
are obtained:}

$\forall \varepsilon >0\exists \delta >0\forall x,y(\left\| x-y\right\|
<\delta \rightarrow \left| p(x)-p(y)\right| <\varepsilon )\longrightarrow
\exists x,y\forall z(p(x)\leq p(z)\leq p(y)).$

An elementary way to complete the proof of Gleason's theorem on the basis of 
\textbf{G5} is in Cooke et al (1984) or Richman and Bridges (1999). Here the
proof is based on a limiting process. Using the continuity of $p$, which has
been proved from \textbf{G1-G4}, the claim that $p$ obtains its minimum and
maximum follows from \textbf{G5}. Using the minimum and maximum points of $p$
one determines the operator $W$, which is the candidate to represent it.
Then one proves that for all $\varepsilon >0$ the proposition $\forall
x(\left| p(x)-(x,Wx)\right| <\varepsilon )$ holds, which completes the proof.

With these observations it is easy to see how Kochen and Specker's theorem
follows from Gleason's theorem. Consider the proposition:

\textbf{F1}. \emph{There is a state }$p$ \emph{such that }$\forall
x(p(x)=0\vee p(x)=1)$

Now, the conjunction of \textbf{F1} with \textbf{G1-G4} is inconsistent,
since the latter imply that $p$ is continuous. Hence, there is a proof of a
contradiction from \textbf{G1-G4 }+ \textbf{F1}. The proof of that
contradiction uses only finitely many cases of the schema \textbf{G3} (since
any proof is finite). If one collects the directions $x\in \mathbb{S}^{2}$
which appear in that proof one gets a finite set of directions on which no
two valued homomorphism exists. Of course this compactness argument does not
yield an explicit set, but it may serve as a incentive to look for one,
which might have been Kochen and Specker's motivation. A similar argument
was explicitly used by Clifton (1993). He simply lifted the vectors which
appear in Bell (1966) simplified 0-1 version of ``Gleason's theorem'' to
obtain a KS theorem. See also Fine and Teller (1978), Pitowsky (1982).

But the argument just presented can be easily generalized to include many
more propositions which contradict Gleason's theorem. Let $\Gamma \subset $ $%
\mathbb{S}^{2}$ be a finite set such that $x\in \Gamma \rightarrow -x\in
\Gamma $. We shall say that $p:\Gamma \rightarrow \mathbb{R}$\emph{\ is a
state on }$\Gamma $ if $p$ satisfies \textbf{G1-G3} for all directions \emph{%
in }$\Gamma $. Now, consider the statement

\textbf{F2} \emph{There is a state} $p$\emph{\ that has exactly }$k$\emph{\
values (}$k\geq 2$\emph{). In other words, }$p$\emph{\ satisfies the
proposition:}

$\mathcal{A}_{k}=\exists x_{1},x_{2,}...x_{k}\bigwedge_{i\neq
j}(p(x_{i})\neq p(x_{j}))\wedge \forall y(p(y)=p(x_{1}))\vee ...\vee
(p(y)=p(x_{k}))$

This contradicts Gleason's theorem since, again by continuity, if $p$ has
two or more values it has infinitely many. Hence, for all $k\geq 2$ there is
a finite set $\Gamma $ which contains elements $x_{1},x_{2,}...x_{k},y$
among others, and such that any state $p$ on $\Gamma $ which assigns $k$
distinct values to $x_{1},x_{2,}...x_{k}$. assigns a different value to $y$.
Also, taking the disjunction $\bigvee_{k=2}^{n}\mathcal{A}_{k}$, we obtain
by the same method that \emph{for each }$n\geq 2$\emph{\ there is a finite
set }$\Gamma _{n}$\emph{\ such that every non constant state }$p$\emph{\ on }%
$\Gamma _{n}$\emph{\ has at least }$n$\emph{\ values}. We shall give below
an explicit construction of $\Gamma _{n}$ in a somewhat more restricted
context.\footnote{%
This result has been used in Breuer (2002) to give an argument against the
``way around'' KS (Meyer 1999, Clifton and Kent 2000, see also, Pitowsky
1983,1985, and Appleby 2002).}

So far we have used only the continuity of $p$, which is proved by \textbf{%
G1-G4}, but Gleason's theorem puts more severe restrictions on states than
continuity. Conceptually, one of the important outcomes of Gleason's theorem
are the uncertainty relations. Casting it in our language it says that any
two non-orthogonal, non-opposite directions cannot both have extreme
probability values (zero or one) unless they are both zero. To see the
finite version consider the opposite statement:

\textbf{F3} \emph{There is a state }$p$ \emph{such that}

$\exists x,y(0<(x,y)<1)\wedge ((p(x)=p(y)=1)\vee (p(x)=1\wedge p(y)=0)\vee
(p(x)=0\wedge p(y)=1))$

Since \textbf{F3 }is false we can prove the following: \emph{Given any }$x,y$%
\emph{\ with }$0<(x,y)<1$\emph{\ there is a finite set }$\Gamma $\emph{\
such that }$x,y\in \Gamma $\emph{, and every state }$p$\emph{\ on }$\Gamma $%
\emph{\ satisfies }$p(x),p(y)\in \{0,1\}\longleftrightarrow p(x)=p(y)=0$.
This is the logical indeterminacy principle (Pitowsky 1998) which has been
proved by an explicit construction, a simplified construction is given
below. Note that this result is stronger than KS since it is constraining
every \emph{probability} distribution on $\Gamma $, and not merely the
``truth values''. It is ``logical'' in the sense that it follows from the
orthogonality relations alone.

We can obtain more dramatic results of this kind, using the fact that by
Gleason's theorem $p(x)=(x,Wx).$ However, recall that this consequence is
derived in the form $\forall \varepsilon >0\forall x(\left|
p(x)-(x,Wx)\right| <\varepsilon )$. We should therefore be careful when
moving to finite subsets. Let us begin with the simple example of a pure
state. If we know that $p(z_{0})=1$ then, by Gleason's theorem, $p(x)=\left|
(z_{0},x)\right| ^{2}$ for all $x$. The statement $(p(z_{0})=1)\wedge
\exists x(p(x)\neq \left| (z_{0},x)\right| ^{2})$ contradicts Gleason's
theorem, but it is refuted by showing that given $x$, and given $\varepsilon
>0$ the condition $\left| p(x)-\left| (z_{0},x)\right| ^{2}\right|
<\varepsilon $ is satisfied. Hence, one cannot expect to be able to force
the relation $p(x)=\left| (z_{0},x)\right| ^{2}$ for each $x$ on a finite
set that contains it. Therefore, consider

\textbf{F4} \emph{There is a state }$p$ \emph{that satisfies}$\
(p(z_{0})=1)\wedge \exists x(\left| p(x)-\left| (z_{0},x)\right| ^{2}\right|
>\varepsilon )$.\emph{for some fixed }$\varepsilon >0$.

The proposition \textbf{F4} clearly contradicts Gleason's theorem. Using our
method we conclude: \emph{For all }$\varepsilon >0$\emph{\ and }$z_{0},x\in $%
\emph{\ }$\mathbb{S}^{2}$\emph{\ there is a finite set of directions }$%
\Gamma $\emph{\ such that }$z_{0}$\emph{\ }$x\in \Gamma $\emph{, and every
state }$p$\emph{\ on }$\Gamma $\emph{\ satisfies: }$p(z_{0})=1\rightarrow $%
\emph{\ }$\left| p(x)-\left| (z_{0},x)\right| ^{2}\right| <\varepsilon $%
\emph{. }Obviously, this is also true for any finite number of directions
beside $x$. The general case, that of a mixture $W$, follows the same
pattern. Here it is not enough to specify the value of $p$ at one point $%
z_{0}$. Rather, five points are needed since, in general, $W$ is a $3\times
3 $, self adjoint, non negative matrix with trace unity. Given these points $%
z_{1},...,z_{5}$, and the values $p(z_{i})=\alpha _{i}$, we find for each $x$
and $\varepsilon >0$ a finite set on which the conditions $p(z_{i})=\alpha
_{i}$ imply $\left| p(x)-(x,Wx)\right| <\varepsilon $. In order construct
this set of directions explicitly one can painstakingly follow the steps of
the constructive proof of Richman and Bridges (1999), and ``lift'' the
vectors in the proof. An alternative to this tedious procedure is presented
in the third section below, where there is a generic algorithmic way to
calculate such graphs (and to demonstrate that Gleason's theorem, and
theorems like it, have a constructive proof).

All these results are easily extendable to any real or complex Hilbert space
of a finite dimension $n\geq 3$, and they are significant for the Bayesian
approach developed in Pitowsky (2003). The results proved here imply that
there are \emph{finite} quantum gambles in which a rational agent is forced
to bet in accordance with the numerical values of quantum probability, or
very near them.

The inverse of these compactness results is the claim that there are very
large subsets $\Omega \subset \mathbb{S}^{2}$ on which 0-1 valued states
exist. The ``size'' of such possible $\Omega $ depends on set-theoretic
assumptions. For example, if the continuum hypothesis is assumed to hold,
there is an $\Omega $ whose intersection with every major circle $C$ in $%
\mathbb{S}^{2}$satisfies $\left| C\setminus \Omega \cap C\right| \leq \aleph
_{0}$. Weaker assumptions lead to ``smaller'' sets (Pitowsky 1983,1985).

\section{Some constructions}

In this section we shall be using rays (one dimensional subspaces) rather
than unit vectors and take states to be defined on them. Given a Hilbert
space $\mathbb{H}$, the assumption $p(\alpha x)=p(x)$ for every scalar $%
\alpha ,\left| \alpha \right| =1$ and every unit vector $x\in \mathbb{H}$
imply that $p$ actually depends on the ray and not on the unit vector we
choose to represent it. Our first aim is to prove the ``logical uncertainty
principle''. The proof here is simpler than Pitowsky (1998) and is based on
the ``lifting'' of the vectors in an argument of Piron (1976).

\begin{theorem}
Let $a$ and $b$ be two non orthogonal rays in a Hilbert space $H$ of finite
dimension $\geq 3$. Then there is a finite set of rays $\Gamma (a,b)$ such
that $a,b\in \Gamma (a,b)$ and such that a state $p$ on $\Gamma (a,b)$
satisfy $p(a),p(b)\in \{0,1\}$ only if $p(a)=p(b)=0$.
\end{theorem}

\begin{proof}
First, consider the three dimensional real space $\mathbb{R}^{3}$. If $z$
and $q$ are two rays in that space there is a unique great circle which they
determine. Let $q^{\prime }$ be the ray orthogonal to both $z$ and $q$ and
let $q^{\prime \prime }$ be the ray orthogonal to both $q$ and $q^{\prime }$%
. Now, consider great circle through $q$ and $q^{\prime }$ ( figure 1).%
\FRAME{ftbpF}{2.8132in}{2.1612in}{0pt}{}{}{ph1s.eps}{\special{language
"Scientific Word";type "GRAPHIC";maintain-aspect-ratio TRUE;display
"USEDEF";valid_file "F";width 2.8132in;height 2.1612in;depth
0pt;original-width 8.3385in;original-height 6.3866in;cropleft "0";croptop
"1";cropright "1";cropbottom "0";filename '../../My
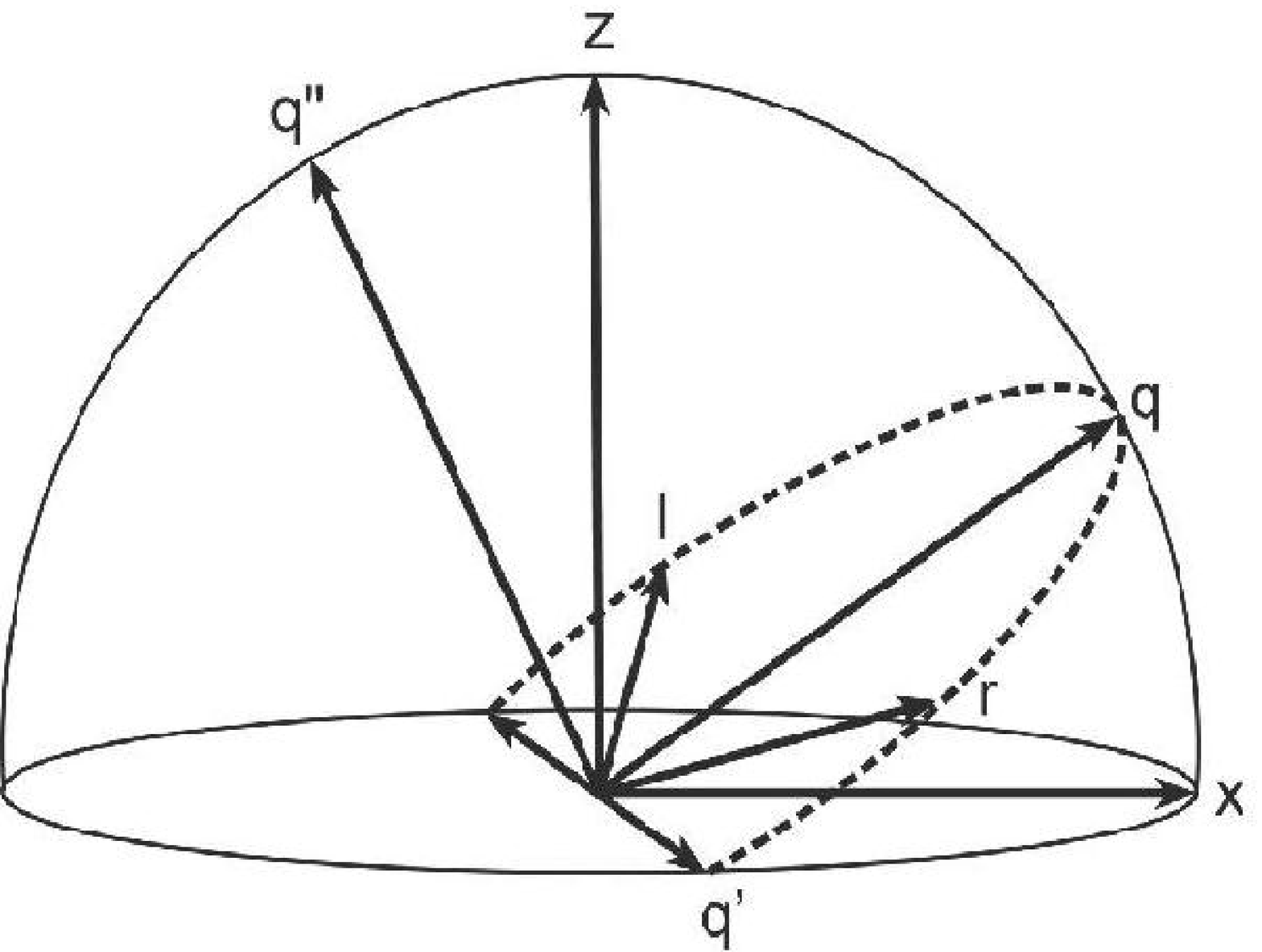';file-properties "XNPEU";}} If $r$ is any ray through
this great circle then $r\perp q^{\prime \prime }$. Let $l$ be the ray
orthogonal to both $r$ and $q^{\prime \prime }$. The orthogonality relations
between $z,q,q^{\prime },q^{\prime \prime },l,r$ is given in the graph $%
G=G(z,q,r)$\ (figure 2).\FRAME{ftbpF}{1.3586in}{2.1612in}{0pt}{}{}{ph2s.eps}{%
\special{language "Scientific Word";type "GRAPHIC";maintain-aspect-ratio
TRUE;display "USEDEF";valid_file "F";width 1.3586in;height 2.1612in;depth
0pt;original-width 4.1684in;original-height 6.6833in;cropleft "0";croptop
"1";cropright "1";cropbottom "0";filename '../../My
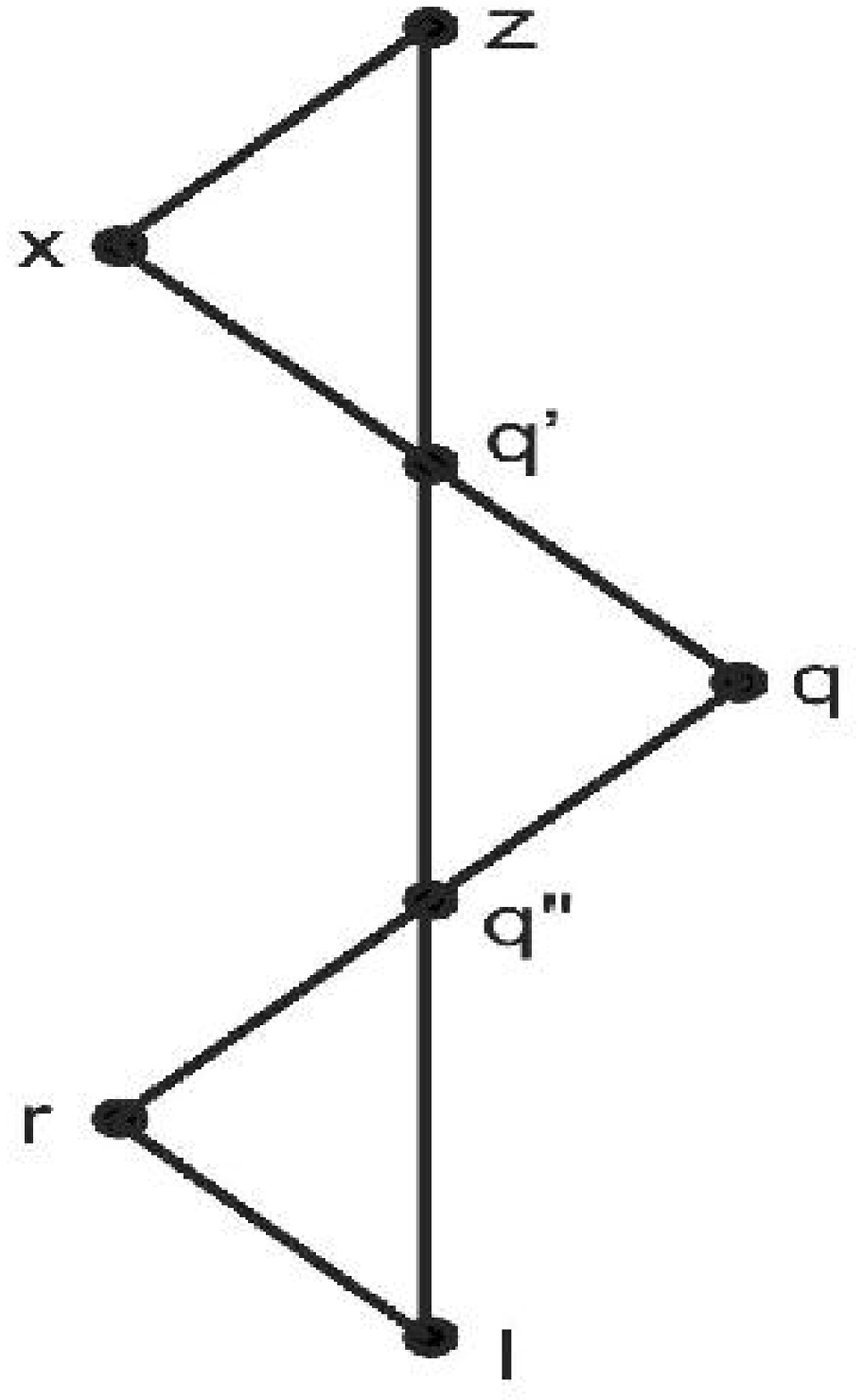';file-properties "XNPEU";}} Subsequently we shall
loosely identify sets of rays with their orthogonality graphs. If $p$ is a
state defined on the rays in the graph $G$ then $p(z)=1$ entails $p(q)\geq
p(r)$. Indeed, $p(q)+p(q^{\prime })+p(q^{\prime \prime
})=p(r)+p(l)+p(q^{\prime \prime })=1$. Also, since $p(z)=1$ we have $%
p(q^{\prime })=0$. Hence $p(q)=p(r)+p(l)\geq p(r)$.

The relation between the points $z,q,$ and $r$ can be best depicted on the
projective plane, where $z$ is taken as the pole of projection (figure 3).%
\FRAME{ftbpF}{2.8755in}{2.1612in}{0pt}{}{}{ph3s.eps}{\special{language
"Scientific Word";type "GRAPHIC";maintain-aspect-ratio TRUE;display
"USEDEF";valid_file "F";width 2.8755in;height 2.1612in;depth
0pt;original-width 8.8894in;original-height 6.6591in;cropleft "0";croptop
"1";cropright "1";cropbottom "0";filename '../../My
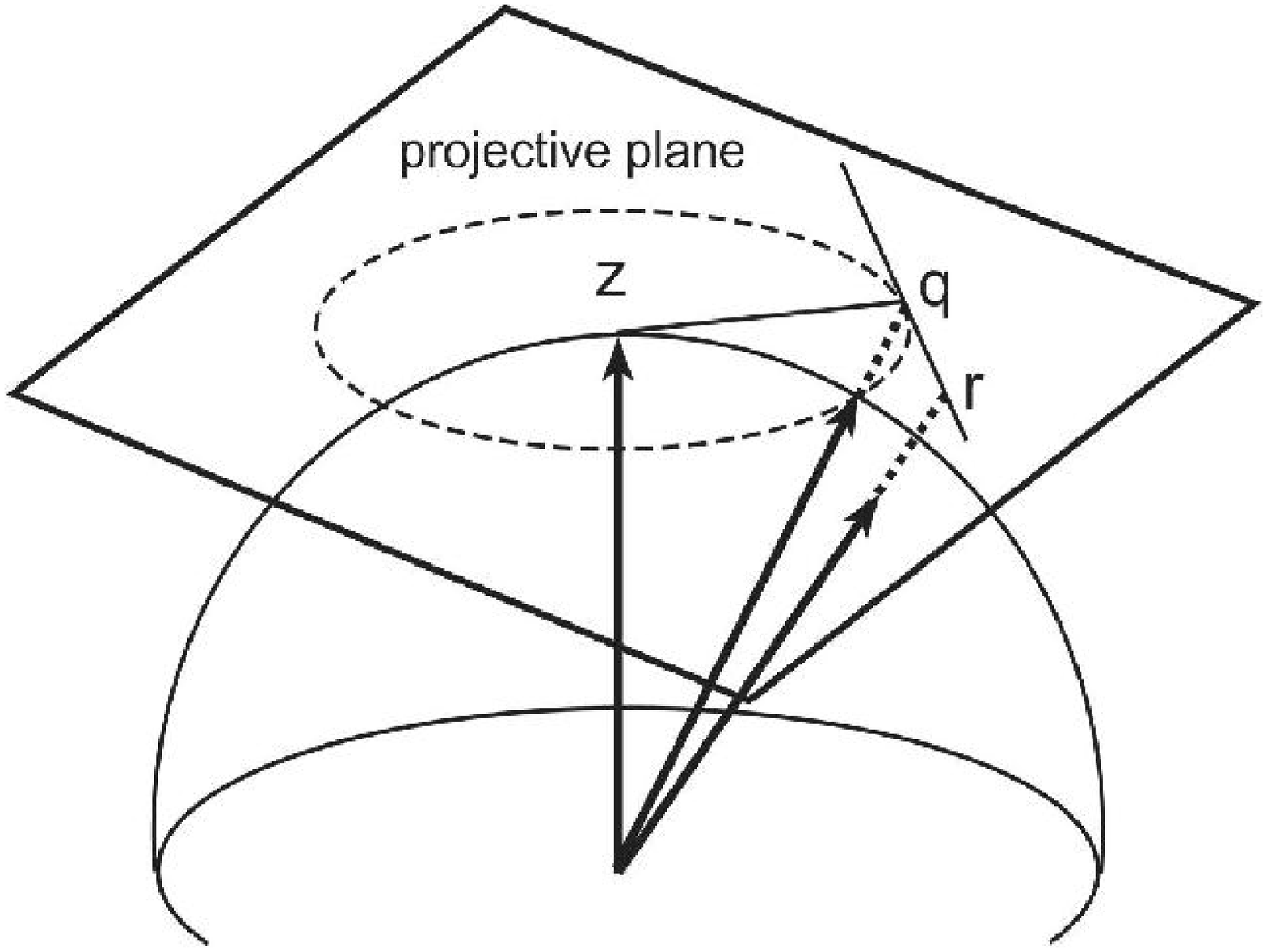';file-properties "XNPEU";}} In the projective plane
great circles appear as straight lines, and latitudes (relative to $z$ as
the pole) appear as concentric circles. In the projective plane $q$ is on a
line through $z$, call this line $\mathcal{L}(z,q)$, and $r$ is on the line
through $q$ which is perpendicular to $\mathcal{L}(z,q)$ at $q$..

Next, consider three points $z,q,r$ which do not necessarily have that
relation. Assume only that $\measuredangle zq<\measuredangle zr$ so that if $%
z$ is the north pole, then $r$ is more to the south than $q$. In this case
we can find a finite sequence of points $q_{1},q_{2},...,q_{m}$ with $q_{1}=q
$ and $q_{m}=r$ and such that $q_{k+1}$ is on the line perpendicular to $%
\mathcal{L}(z,q_{k})$ for $k=1,2,...,m-1$. A case with $m=5$ is considered
in figure 4\FRAME{ftbpF}{2.2511in}{2.1612in}{0pt}{}{}{ph4s.eps}{\special%
{language "Scientific Word";type "GRAPHIC";maintain-aspect-ratio
TRUE;display "USEDEF";valid_file "F";width 2.2511in;height 2.1612in;depth
0pt;original-width 6.9444in;original-height 6.6642in;cropleft "0";croptop
"1";cropright "1";cropbottom "0";filename '../../My
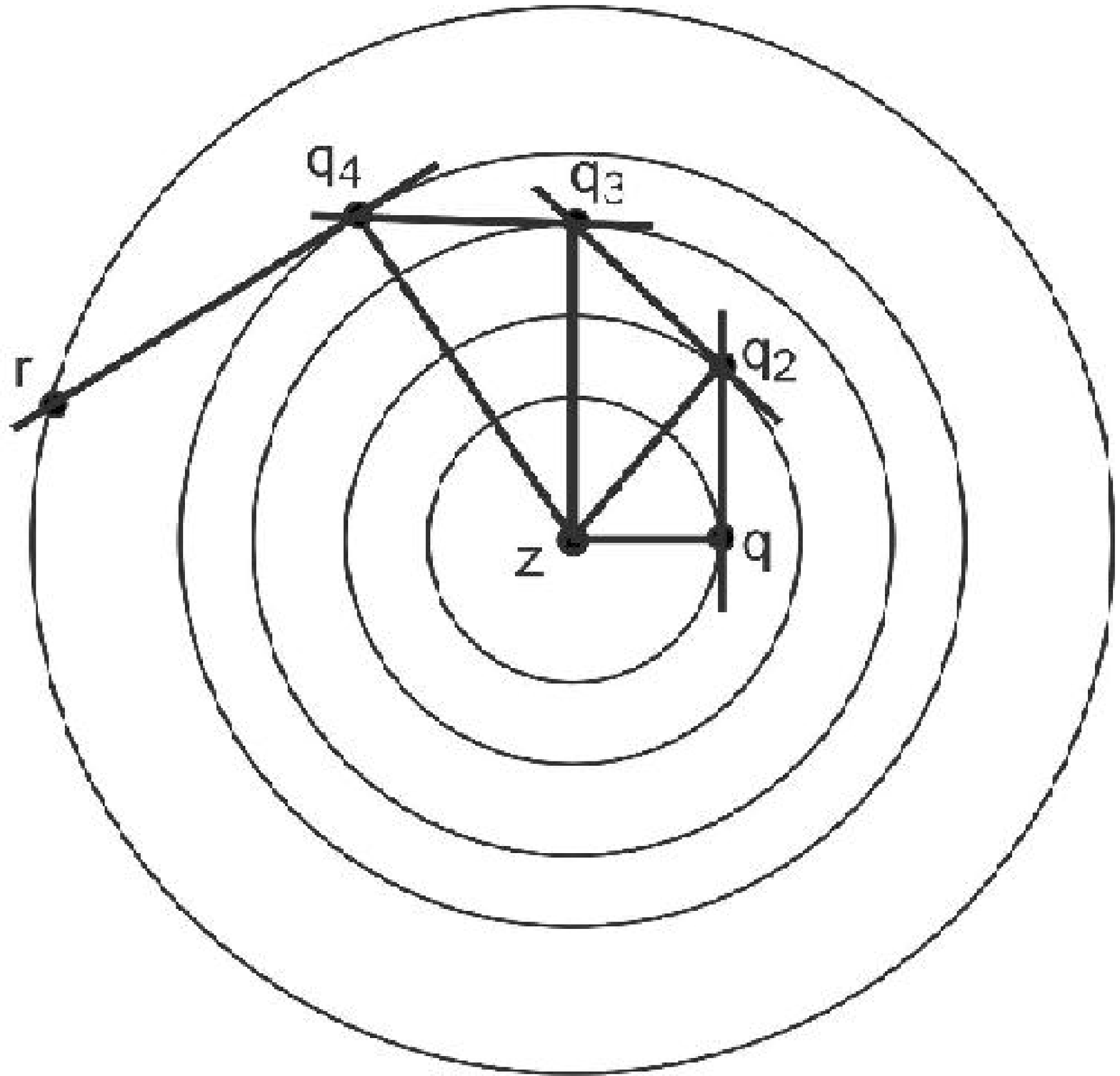';file-properties "XNPEU";}}. The number $m$ of
intermediate points depends on the difference $\measuredangle
zr-\measuredangle zq$ and on the respective longitude of $q$ and $r$.

Given this set of vectors we can construct for each $k$ a graph $%
G_{k}=G(z,q_{k},q_{k+1})$ in which $z,q_{k},q_{k+1}$ play the role of $z,q,r$
respectively (note, $z$ is the same throughout). Let $G^{\prime
}(z,q,r)=\bigcup_{k=1}^{m-1}G_{k}$ then any state on $G^{\prime }$ that
satisfy $p(z)=1$ also satisfy 
\begin{equation*}
p(q)=p(q_{1})\geq p(q_{2})\geq ...\geq p(q_{n})=p(r)
\end{equation*}

To finish the proof for $\mathbb{R}^{3}$ let $a,b$ be two non orthogonal
rays. We can always choose a sequence of rays $c_{1},c_{2},...,c_{n}$ such
that $\measuredangle ab>\measuredangle bc_{1}>\measuredangle
c_{1}c_{2}>...>\measuredangle c_{n-1}c_{n}$ and such that $a\perp c_{n}$. A
case with $n=3$ is depicted in figure 5\FRAME{ftbpF}{2.9412in}{2.1612in}{0pt%
}{}{}{ph5s.eps}{\special{language "Scientific Word";type
"GRAPHIC";maintain-aspect-ratio TRUE;display "USEDEF";valid_file "F";width
2.9412in;height 2.1612in;depth 0pt;original-width 8.3385in;original-height
6.1073in;cropleft "0";croptop "1";cropright "1";cropbottom "0";filename
'../../My 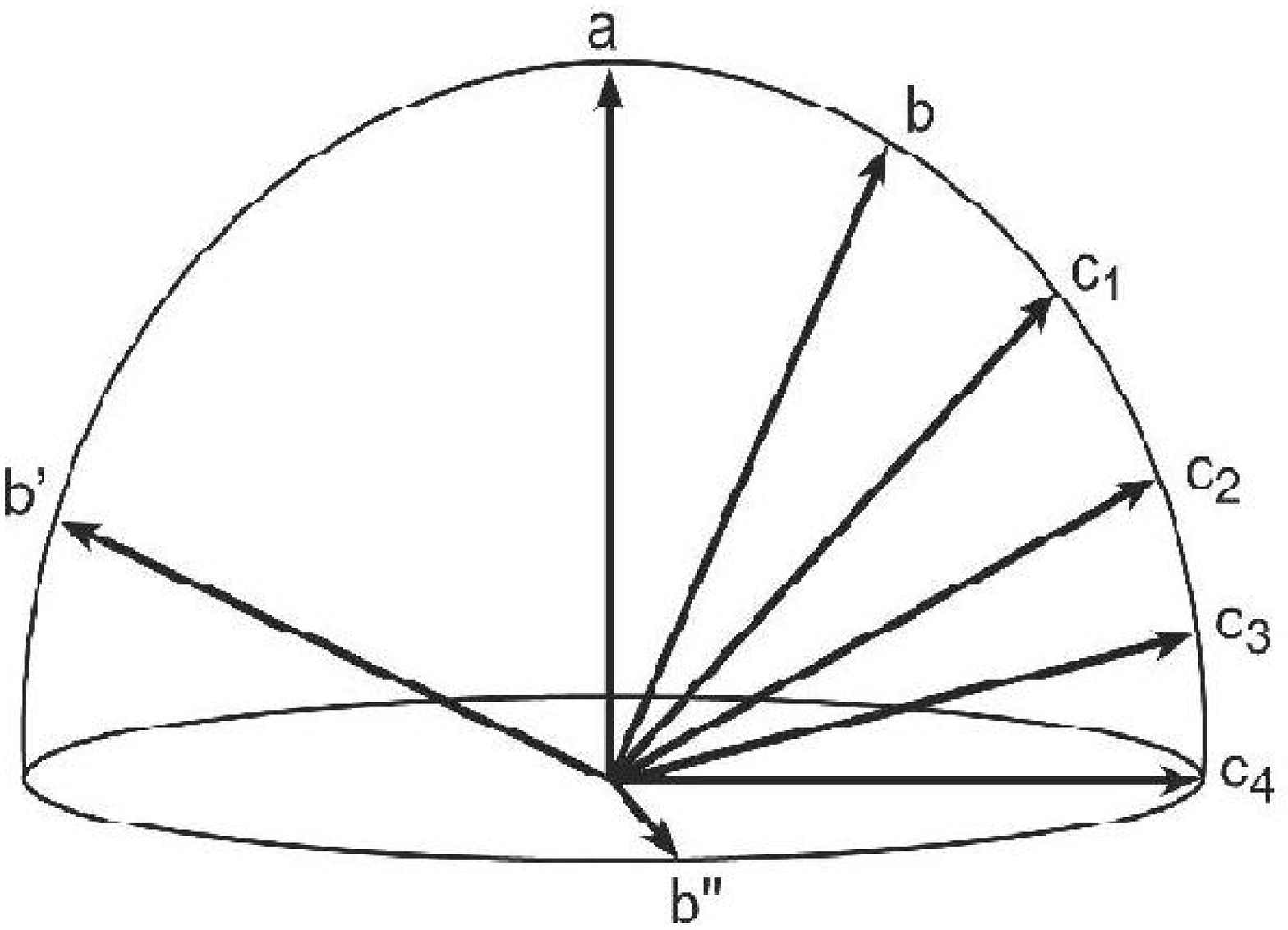';file-properties "XNPEU";}}.Consider $b$ as
a pole (projection point) and construct a graph $G_{0}^{\prime }=$ $%
G^{\prime }(b,c_{1},a)$ which is like $G^{\prime }(z,q,r)$ with $b,c_{1},a$
play the role of $z,q,r$ respectively. If $p$ is a state on the rays in this
graph with $p(b)=1$, then $p(c_{1})\geq p(a)$. Consider $c_{1}$ as a pole
and construct the graph $G_{1}^{\prime }=G^{\prime }(c_{1},c_{2},b)$, which
is like $G^{\prime }(z,q,r)$, with $c_{1},c_{2},b$ play the role of $z,q,r$
respectively. For a probability function $p$ on $G_{1}^{\prime }$ which
satisfy $\ p(c_{1})=1$ we get $p(c_{2})\geq p(b)$. Now construct the graph $%
G_{2}^{\prime }$ with $c_{2}$ as pole and $c_{3},c_{1}$ play the role of $q,r
$ respectively, and so on. Suppose that $p$ is a state on $G^{\prime \prime
}(a,b)=\bigcup_{j=0}^{n-1}G_{j}^{\prime }$ the union of all these graphs. We
shall show that if $p(a)=1$ then $p(b)<1$. Assume, by negation, that $%
p(a)=p(b)=1$ then, by construction $p(c_{1})\geq p(a)$ so that $p(c_{1})=1$.
But then $p(c_{2})\geq p(b)$ so that $p(c_{2})=1$, and $p(c_{3})\geq p(c_{1})
$ so that $p(c_{3})=1$, and so on and we finally obtain $p(c_{n})=1$. This
is a contradiction since $a\perp c_{n}$. Hence $p(b)<1$.

Now, consider the ray $b^{\prime }$ which is orthogonal to $b$ in the plane
spanned by $a$ and $b,$ and let $b^{\prime \prime }$ be the ray orthogonal
to both $b,b^{\prime }$. Repeat the construction of the graph with $%
b^{\prime }$ instead of $b$ to obtain the graph $G^{\prime \prime
}(a,b^{\prime })$, and add this to the previous graph. Let$\ p$ is a state
on the graph $G^{\prime \prime }(a,b)\cup G^{\prime \prime }(a,b^{\prime })$
with $p(a)=1$, then $f(b)<1$ and $f(b`)<1.$ But then we also have $p(b)>0$.
Otherwise, $p(b)=0$ together with $p(b^{\prime \prime })=0$ (as $b^{\prime
\prime }\perp a$ and $p(a)=1$) entail $p(b^{\prime })=1$, contradiction.
Hence $p(a)=1$ entails $0<p(b)<1$. Inverting the roles of $a$ and $b$ we
construct a graph 
\begin{equation*}
\Gamma (a,b)=G^{\prime \prime }(a,b)\cup G^{\prime \prime }(a,b^{\prime
})\cup G^{\prime \prime }(b,a)\cup G^{\prime \prime }(b,a^{\prime })
\end{equation*}

With $a^{\prime }$ a vector orthogonal to $a$ in the plane spanned by $a$
and $b$. Let $p$ be a state on $\Gamma (a,b)$. If$\ p(b)=1$ then $0<p(a)<1$,
and if$\ p(a)=1$ then $0<p(b)<1$. Therefore $\Gamma (a,b)$ is the required
set of rays in $\mathbb{R}^{3}$.

In the general case of a finite dimensional Hilbert space $\mathbb{H}$ we do
the following: Given rays $a,b$ in $\mathbb{H}$ we consider them first as
rays in a three dimensional subspace $\mathbb{H}^{\prime }$of $\mathbb{H}$
and complete the construction there. Then we add to the finite set of rays
in $\mathbb{H}^{\prime }$additional $\dim \mathbb{H}-3$ orthogonal rays in
the orthocomplement of $\mathbb{H}^{\prime }$. This completes the proof.
\end{proof}

The the above construction entails that pure states should have strictly
monotone behavior

\begin{lemma}
Given a ray $z$ in a Hilbert space $\mathbb{H}$ of a finite dimension $\geq
3 $, and rays $a$ and $b$ such that $0<\measuredangle (a,z)<\measuredangle
(b,z)$ then there is a finite set of rays $D(z,a,b)$, which contains $z,a$,
and $b$, such that every state $p$ on $D(z,a,b)$ for which $p(z)=1$ also
satisfies $p(a)>p(b)$.
\end{lemma}

\begin{proof}
Consider first $\mathbb{R}^{3}$. Given a ray $z$, let $q$ be any ray
different from $z$ and not orthogonal to it. Consider once more the vectors
in figute 1 and thier orthogonality graph $G=G(z,q,r)$ in figure 2. Now
denote 
\begin{equation*}
D_{1}(z,q,r)=G(z,q,r)\cup \Gamma (z,l)
\end{equation*}
where $\Gamma (z,l)$ is the set of rays in theorem 2 with $z,l$ fulfill the
role of $a,b$. The situation is depicted in figure 6. \FRAME{ftbpF}{1.3586in%
}{2.1612in}{0pt}{}{}{ph6s.eps}{\special{language "Scientific Word";type
"GRAPHIC";maintain-aspect-ratio TRUE;display "USEDEF";valid_file "F";width
1.3586in;height 2.1612in;depth 0pt;original-width 4.1684in;original-height
6.6833in;cropleft "0";croptop "1";cropright "1";cropbottom "0";filename
'../../My 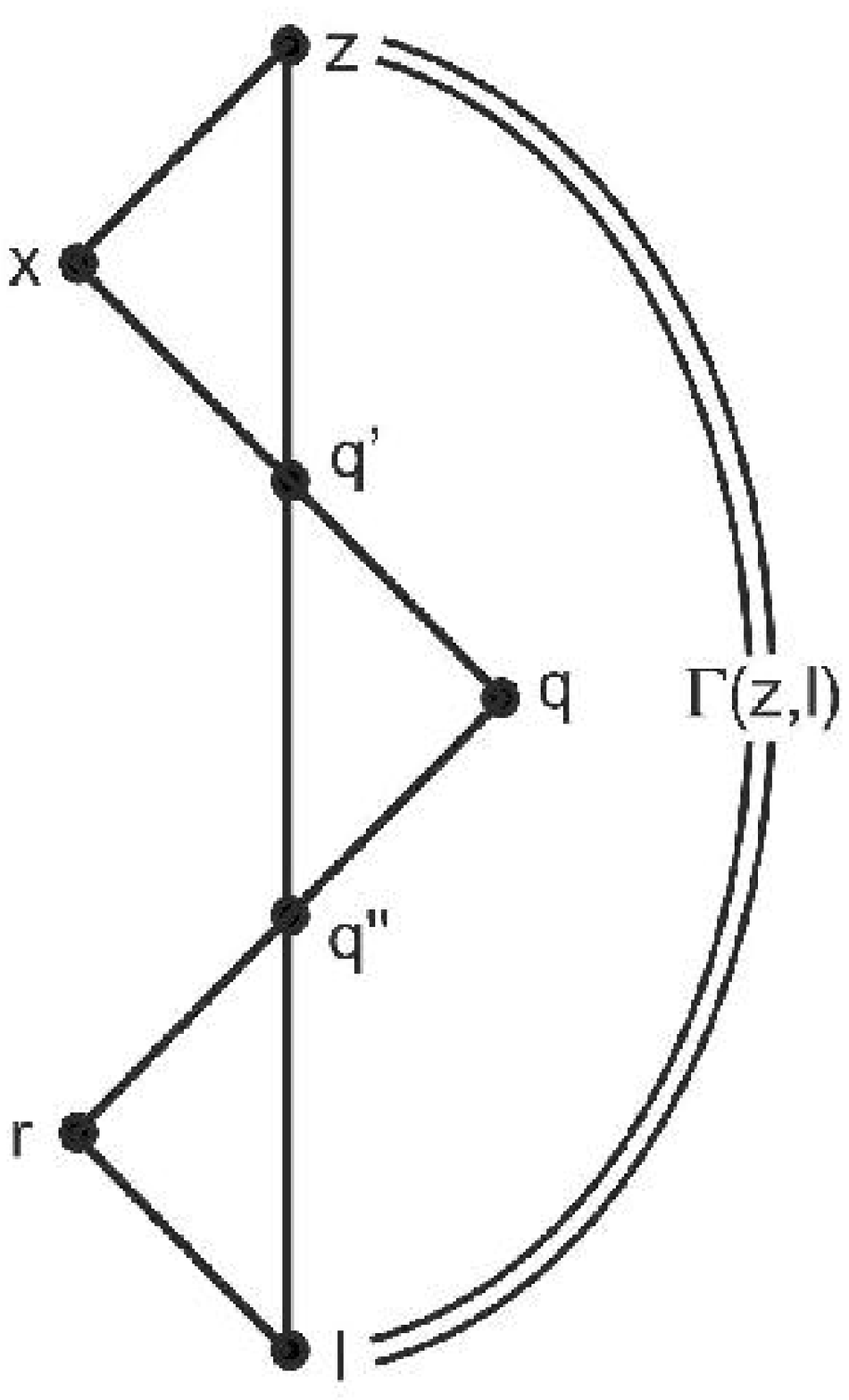';file-properties "XNPEU";}}If $p$ is a
probability function on $D_{1}(z,q,r)$ with $p(z)=1$ then, by construction, $%
p(q)=p(r)+p(l)$, and also $0<p(l)<1$, hence $p(q)>p(r)$. Since $%
\measuredangle (a,z)<\measuredangle (b,z)$ we can find a sequence $q_{0}$, $%
q_{1}$, ...,$q_{m}$ such that $q_{0}=a$ and $q_{m}=b$ and for all $k=1,2,...m
$ the ray $q_{k}$ is on the great circle through $q_{k-1}$ and $%
q_{k-1}^{\prime }$, the ray orthogonal to $z$ and $q_{k-1}$. Putting 
\begin{equation*}
D(z,a,b)=\bigcup_{k=1}^{m}D_{1}(z,q_{k-1,}q_{k})
\end{equation*}
We get that if $p(z)=1$ then 
\begin{equation*}
p(a)=p(q_{0})>p(q_{1})>...>p(q_{m})=p(b)
\end{equation*}
If $\mathbb{H}$ is a Hilbert space $3\leq \dim \mathbb{H}<\infty $, complete
the construction first on the three dimensional space $\mathbb{H}_{1}$
spanned by $z$, $a$, and $b$ (in case they all lie in the same plane form $%
\mathbb{H}$ $_{1}$ by adding any ray orthogonal to them). Subsequently add a
set of orthogonal rays in $\mathbb{H}_{1}^{\bot }$ to complete the
construction.
\end{proof}

An immediate consequence of this theorem is the construction of a finite
sets of rays on which states must take many values:

\begin{corollary}
Given a ray $z$ in a Hilbert space $\mathbb{H}$ of a finite dimension $\geq
3 $, and an integer $k$, there is a finite set of rays $\Lambda _{k}(z)$
such that every state $p$ on $\Lambda _{k}(z)$ for which $p(z)=1$ has at
least $k$ distinct values.
\end{corollary}

\section{The Effectiveness of Gleason's Theorem}

Recently, there has been an interesting discussion on the question whether
Gleason's theorem has a proof which is acceptable by the standards of
constructive mathematics (Hellman 1993, Billinge 1997). The discussion
culminated when Richman and Bridges (1999) gave a constructive formulation
and proof of the theorem. Our aim is to give a (much shorter) proof of the
conditional statement: \emph{If Gleason's theorem is true than it must have
an effective proof.}

The result follows from a generic sequence of approximations of states on
finite sets. As will become clear subsequently the approximations in
question are determined by an algorithm. Again, we shall consider the three
dimensional case but the results generalize immediately. To simplify matters
we shall work with general \emph{frame functions}, not (positive normalized)
states. This means that we replace \textbf{G1} and \textbf{G3} respectively
by the axioms:

\textbf{G'1 }\emph{There is a real constant }$\gamma $ \emph{such that} $%
\forall xp(x)\geq \gamma $.

\textbf{G'3 }\emph{There is a real constant }$\delta $\emph{\ such that for
each orthonormal triple }$x,y,z\in \mathbb{S}^{2}$:\emph{\ }$%
p(x)+p(y)+p(z)=\delta $

Let $e_{1}=(1,0,0)$, $e_{2}=(0,1,0)$ and $e_{3}=(0,0,1)$ be the standard
basis in $\mathbb{R}^{3}$ (or $\mathbb{C}^{3}$) and $b_{ij}=\frac{1}{\sqrt{2}%
}(e_{i}+e_{j})$, $1\leq i<j\leq 3$.

Denote by $\mathcal{F}$ the set of all functions $p:\mathbb{S}%
^{2}\rightarrow \mathbb{R}$ that satisfy \textbf{G'1}, \textbf{G2} and 
\textbf{G'3} and let $\mathcal{F}_{0}=\{p\in \mathcal{F};\
p(e_{i})=p(b_{ij})=0,\ i=1,2,3,\ 1\leq i<j\leq 3\}$. Then

\begin{lemma}
The following statements are equivalent:

(a) If $p\in \mathcal{F}$ there is a self adjoint operator $W$ in $\mathbb{R}%
^{3}$ such that $p(x)=(x,Wx)$.

(b) Every element $p\in \mathcal{F}_{0}$ vanishes identically.
\end{lemma}

\begin{proof}
If $p\in \mathcal{F}_{0}$ then by (a) we have $p(x)=(x,Wx)$. But then $%
(e_{i},We_{i})=0$ and $(b_{ij},Wb_{ij})=0$. The latter equation implies that 
$(e_{i},We_{j})=0$ for $1\leq i<j\leq 3$. Hence $W=0$ and therefore $p=0$.

Conversely, assume (b) holds, let $p\in \mathcal{F}$ and let $W$ be the
symmetric matrix which satisfies the equations $p(e_{i})=(e_{i},We_{i})$ and 
$p(b_{ij})=(b_{ij},Wb_{ij})$ for $i=1,2,3,\ 1\leq i<j\leq 3$. Denote $%
p_{0}(x)=p(x)-(x,Wx)$. Then $p_{0}\in \mathcal{F}$ and $%
p_{0}(e_{i})=p_{0}(b_{ij})=0,$ and therefore $p_{0}\in \mathcal{F}_{0}$. By
(b) $p_{0}=0$ and thus $p(x)=(x,Wx)$.
\end{proof}

By Gleason's theorem (a) is true, so we can take (b) as our formulation of
this theorem. In the following we shall work with the first order
formalization of the field of real numbers, the theory of real closed
fields, denoted by $\mathbf{R}$, or the formalization of the field of
complex numbers, the theory of algebraically closed fields (with zero
characteristic), denoted by $\mathbf{C}$. In both theories there is an
effective elimination of quantifiers. This means that there is a (known)
algorithm which, given any well formed formula as input, produces as output
an equivalent formula without the quantifiers $\forall $, $\exists $.
Consequently the theories are decidable: there is an algorithmic method to
prove every true proposition in them\footnote{%
For details on these theories see Shoenfeld (1967). For recent results on
quantifier elimination see Basu (1999)}. Also, we shall denote by $\mathbf{R}%
_{-}$ the section of $\mathbf{R}$ without multiplication (including only the
addition operation and the inequality relation).

As a result of the elimination of quantifiers every definable set (without
parameters) in $\mathbf{R}$ is a finite Boolean combination of sets of $n$%
-tuples, each defined by a rational polynomial inequality (or equality).
Thus, for example, $\mathbb{S}^{2}$ is the set $\{(\alpha _{1},\alpha
_{2},\alpha _{3});\alpha _{1}^{2}+\alpha _{2}^{2}+\alpha _{3}^{2}=1\}$ and
therefore definable in $\mathbf{R}$. Similarly $\mathbb{S}_{+}^{2}$, ``the
northern hemisphere'', which is like $\mathbb{S}^{2}$ with the additional
condition $(\alpha _{3}>0)\vee (\alpha _{3}=0\wedge \alpha _{1}>0)\vee
(\alpha _{3}=0\wedge \alpha _{1}=0\wedge \alpha _{2}=1)$, is also definable
in $\mathbf{R}$. Similar remarks apply to $\mathbf{C}$ and the unit sphere
in the complex, finite dimensional Hilbert spaces.

We shall consider the general set-up where $X$ is some definable set of $n$%
-tuples, and $Y$\ is a definable set of $m$-tuples from $X$. We let $%
\mathcal{F}_{0}$ be the set of real-valued functions $p$ on $X$, such that $%
\sum_{i=1}^{m}p(x_{i})=0$ for all $y=(x_{1},...,x_{m})\in Y$. Our first aim
is to prove the following: \emph{If every bounded function of} $\mathcal{F}%
_{0}$ \emph{vanishes identically in every model of} $\mathbf{R}$\emph{\ (or }%
$\mathbf{C}$ \emph{) then this fact admits an effective proof}.

The connection with Gleason's theorem is as follows: We let $X=\mathbb{S}%
_{+}^{2}$, and $Y$ is the set of triples from $X$ given by $Y=B\cup
\{(e_{i},e_{i},e_{i});i=1,2,3\}\cup \{(b_{ij},b_{ij},b_{ij});1\leq i<j\leq
3\}$. Here $B$ is the set of orthogonal triples from $\mathbb{S}_{+}^{2}$
and $e_{i}$, $b_{ij}$ as in lemma 2. $B$ is clearly definable since the
inner product of $x_{1}=(\alpha _{1},\alpha _{2},\alpha _{3})$, $%
x_{2}=(\beta _{1},\beta _{2},\beta _{3})$ is a polynomial in the $\alpha
_{i} $'s and $\beta _{i}$'s. Also, the $b_{ij}$'s are given by a simple
formula (for example, $b_{12}=(\alpha ,\alpha ,0)$ with $2\alpha ^{2}-1=0$).
In sum, $Y$ is definable. In this case $\mathcal{F}_{0}$ is the set of
real-valued functions $p$ which satisfy $p(x_{1})+p(x_{2})+p(x_{3})=0$ for
all $(x_{1},x_{2},x_{3})\in Y$. By lemma 2 the conclusion $p=0$ for all $%
p\in \mathcal{F}_{0}$ is equivalent to Gleason's theorem.

Now for a natural number $k>1$ let

\begin{equation*}
\mathcal{S}_{k}(n)=\{S\subset \{0,\ldots ,n\}^{m}:\mathbf{R}_{-}\vdash
(\forall \alpha _{0})\ldots (\forall \alpha _{n})\left( \bigwedge_{s\in
S}(\sum_{i=1}^{m}\alpha _{s(i)}=0)\rightarrow \bigvee_{j=1}^{n}(k|\alpha
_{0}|\leq |\alpha _{j}|)\right) \}.
\end{equation*}

Let $\mathbf{T}$ stand for either $\mathbf{R}$ or $\mathbf{C}$ then we have:

\begin{lemma}
Let $X$ be a definable set of $n$-tuples, $Y$ a definable set of $m$-tuples
from $X$, and $\mathcal{F}_{0}$ the set of real-valued functions $p$ on $X$,
such that $\sum_{i=1}^{m}p(x_{i})=0$ for all $y=(x_{1},...,x_{m})\in Y$.
Then the following are equivalent

\begin{enumerate}
\item  Every bounded function in $\mathcal{F}_{0}$ vanishes in every model
of $\mathbf{T}$.

\item  For some $n$, $\mathbf{T}\vdash (\forall x_{0})(\exists x_{1})\ldots
(\exists x_{n})\bigvee_{S\in \mathcal{S}_{k}(n)}\bigwedge_{s\in S}(y_{s}\in
Y)$ where $y_{s}=(x_{s(1)},\ldots ,x_{s(m)})$.
\end{enumerate}
\end{lemma}

\begin{proof}
Let $p\in \mathcal{F}_{0}$. If (2) holds for $n$, then there is $S\in 
\mathcal{S}_{k}(n)$ such that $\sum_{i=1}^{m}p(x_{s(i)})=0$ for all $s\in S.$
By the definition of $\mathcal{S}_{k}(n)$ we have $\left| p(x_{0})\right|
\leq \frac{1}{k}\left| p(x_{j})\right| $ for some $1\leq j\leq n$. But by
(2) for each $x_{0}\in X$ we can find such $x_{j}\in X$. We conclude
therefore that if $\beta >0$ bounds $p$, so does $(1/k)\beta $. Thus
repeating the argument we have proved $p=0$.

Conversely, assume that (2) fails. Note that the formulas

\begin{equation*}
\phi _{n}(x_{0})=(\exists x_{1})\ldots (\exists x_{n})\bigvee_{S\in \mathcal{%
S}_{k}(n)}\bigwedge_{s\in S}((x_{s(1)},\ldots ,x_{s(m)})\in Y)
\end{equation*}
define an increasing chain of sets $W_{k}(n)=\{a;\phi _{n}(a)\}$, so their
complements $\overline{W_{k}}(n)=\{a;\lnot \phi _{n}(a)\}$ form a decreasing
chain, and by assumption no element is empty. By the compactness theorem (or
G\"{o}del's completeness theorem), there exists a model $\mathcal{M}$ of $%
\mathbf{T}$ and $a_{0}\in \left| \mathcal{M}\right| $ such that $a_{0}\in 
\overline{W}_{k}(n)$ for all$\ n$.

We now show that some $p\in \mathcal{F}_{0}(\mathcal{M})$ is non zero. For
each $a\in X(\mathcal{M})$, let $c_{a}$ be a new (and different) constant
symbol; let $\overline{\mathbf{R}}$ be the theory consisting of the axioms
of $\mathbf{R}_{-}$ and for each $(a_{1},\ldots ,a_{m})\in Y(\mathcal{M})$
the axiom $\sum_{i=1}^{m}c_{a_{i}}=0$, and for each $a\in X(\mathcal{M})$
the axiom $|c_{a}|<k|c_{a_{0}}|$. Then $\overline{\mathbf{R}}$ is
consistent, so it also has a model. Hence, there exists a function $g$ on $X(%
\mathcal{M})$ into an ordered Abelian group $(B,<)$ with $\sum g(a_{i})=0$
for all $(a_{1},...,a_{m})\in Y(\mathcal{M})$ and with $|g(a)|<k\left|
g(a_{0})\right| $ for all $a\in X(\mathcal{M})$. In particular, $%
b_{0}:=|g(a_{0})|>0$. Also, there exists a unique homomorphism $%
H:B\rightarrow \mathbf{R}$ with $H(z)\geq 0$ whenever $z\geq 0$, and $%
H(b_{0})=1$ . Now $p=H\circ g$ is a nonzero element of $\mathcal{F}_{0}(%
\mathcal{M}),$ and therefore (1) fails.
\end{proof}

This establishes the existence of a constructive proof for Gleason's theorem
in complex Hilbert spaces. In this case, the condition ``every bounded
function in $\mathcal{F}_{0}$ vanishes'' holds in every model of $\mathbf{C}$%
, iff it holds in $\mathbb{C}$\footnote{%
Knowing the theorem for $\mathbb{C}$ automaticaly imply it for every model,
since $\mathbb{C}$ is universal for countable models of $\mathbf{C}$;
however, $\mathbb{R}$ does not have that property.}. This means that
condition (2) of the theorem states that Gleason's theorem is true iff $%
\forall x_{0}\phi _{n}(x_{0})$ is provable in $\mathbf{C}$ for some $n$. But 
$\mathbf{C}$ has an effective decision procedure.

In the real case there is some complication because we do not assume
Gleaason's theorem for every model , but only for the standard model; though
a posteriori it will follow for every model. Therefore, we have to bridge
the gap between ``all models of $\mathbf{R}$'' and the standard model $%
\mathbb{R}$ for which Gleason's theorem is known to be true. From now on we
specialize to the concrete $Y$ relevant to Gleason's theorem. More
specifically, we shall take $X$ to be the real projective plane, using its
identification with the northern hemisphere $\mathbb{S}_{+}^{2}$ of the
previous section (Figure 3). With this identification (and with the point at
infinity added) $X$ is compact and $Y$ is closed. It follows that the sets $%
W_{k}(n)$, being projections of closed sets, are also closed sets, and so
the sets ,${\overline{W}}_{k}(n)$ are open; but all we need is their
measurability with respect to the uniform measure on the sphere, and the
measurability of their intersections with major circles with respect to the
uniform measure on the circle. Since $W_{k}(n)$ are definable sets, the
subsets of the circle that will be mentioned below are finite unions of
segments (i.e. arcs) and their measure is elementary.

\begin{theorem}
The conditions below, referring to the model $\mathbb{R}$, are equivalent.

\begin{enumerate}
\item  Every bounded function in $\mathcal{F}_{0}$ vanishes.

\item  For some $n$, $\mathbf{T}\vdash (\forall x_{0})(\exists x_{1})\ldots
(\exists x_{n})\bigvee_{S\in {\mathcal{S}_{2}(n)}}\bigwedge_{s\in
S}(y_{s}\in Y)$ where $y_{s}=(x_{s(1)},\ldots ,x_{s(m)})$. (that is, for
some $n$, ${\overline{W}}_{2}(n)=\emptyset $)

\item  For some $n$, \ for every great circle $C$, the set $C\cap {W}_{4}(n)$
has normalized measure $>1/2$ in $C$.

\item  $\cap _{n\geq 1}{\overline{W}}_{8}(n)=\emptyset $.
\end{enumerate}
\end{theorem}

Condition (2), while phrased for $\mathbb{R}$, is a single elementary
statement; thus if true it holds in every real closed field, and this fact
admits an elementary proof. So in the special case of Gleason's theorem, we
can replace ``in every model of $\mathbf{R}$'' by ``in $\mathbb{R}$.''.

\begin{proof}
We will use the following kind of Fubini principle: Let $C_{z}$ be the great
circle orthogonal to the point $z$ on the sphere. For $C=C_{z}$ let $\mu
_{C} $ be the normalized uniform measure on $C_{z}$; and say that a property 
$P$ holds ``for a majority of points'' if $\mu _{C}(\{x:P(x)\})\geq 1/2$.
Let $C$ be a fixed great circle and suppose that a majority of points $z\in
C $ are such that for a majority of points $q\in C_{z}$, the property $P(q)$
holds. Then, assuming $\{x:P(x)\}$ is measurable on the sphere, it has an
area $\geq A$, where $A$ is the area of the unit sphere above the 45th
northern latitude. The reason is that the region $\{x:P(x)\}$ takes up $\geq
1/4$ of the area of a Mercator map of the sphere; and the region above the
45 latitude has the greatest distortion under this map. Moving to the
projective plane $X$, we can say that $\{x:P(x)\}$ takes an area $\geq 1/2$
of $X$, when we use the area measure induced on $X$ by the Mercator map.

\begin{claim}
If $z\in {\overline{W}}_{k}(2n+2)$, then a majority of $x\in C_{z}$ lie in ${%
\overline{W}}_{2k}(n)$.
\end{claim}

Indeed, in the projective plane model, a point $x$ on a given circle $%
C=C_{z} $, has a unique point $ort(x)$ orthogonal to it on $C_{z}$. Suppose
the claim fails. So let $z\in {\overline{W}}_{k}(2n+2)$ and assume that a
set of points of $C_{z}$ of measure $>1/2$ lies in $W_{2k}(n)$. Since the
function $ort$ is a measure-preserving bijection on $C_{z}$, the set $%
\{y:ort(y)\in W_{2k}(n)\}$ also has measure $>1/2$. So these two sets
intersect; thus there exist $x,y\in C_{z}$ such that $x\bot y$ and $x,y\in
W_{2k}(n)$. But as $x,y,z$ form an orthogonal triple, $(x,y,z)\in Y$, so it
is easy to see that $z\in W_{k}(2n+2)$, a contradiction which proves the
claim.

\textbf{(4) implies (3)}: Assuming (4), there exists $n_{4}$ such that ${W}%
_{8}(n_{4})$ has measure $>A$, that is, Mercator measure $>1/4$. Let $%
n_{3}=2n_{4}+2$. Let $C$ be a great circle. Suppose, by negation, that a
majority of points of $C$ lie in ${\overline{W}}_{4}(n_{3})$. Then, by the
claim, for each of these points $z\in {\overline{W}}_{4}(n_{3})\cap C$, a
majority of points $x$ of $C_{z}$ must lie in ${\overline{W}}_{8}(n_{4})$.
So this set has (Mercator) area $\geq 1/4$. Moving to the projective plane
we get a contradiction. Thus (3) holds.

\textbf{(3) implies (2)}: Assume (3) holds for $n=n_{3}$, and let $%
n_{2}=2n_{3}+2$. If $y\in {\overline{W}}_{2}(n_{2})\neq \phi $, let $C_{y}$
be the orthogonal great circle. Then by the claim a majority of points on $%
C_{y}$ lie in ${\overline{W}}_{4}(n_{3})$, contradicting (3).

\textbf{(2) implies (1)}: The same as in lemma 3.

\textbf{(1) implies (4)}: As soon as there is $a_{0}\in \cap _{n\geq 1}{%
\overline{W}}_{8}(n)$, the proof of lemma 3 works and provides a
non-vanishing bounded function in $\mathcal{F}_{0}$.
\end{proof}

The theorem provides an effective procedures to prove $p=0$ in case every $%
p\in \mathcal{F}_{0}$ vanishes. The reason is that $\mathbf{R}$ (and $%
\mathbf{R}_{-}$) is decidable, which means that there is a computer program
which proves evry true propositions of $\mathbf{R}$, in particular the
proposition $(\forall x_{0})\phi _{n}(x_{0})$ for a suitable integer $n$.
Any apparent ineffectiveness in the proof of theorem 3 is in some sense
self-eliminating, as a posteriori one knows the existence of an effective
proof too. Nevertheless, it may be worth remarking that all the sets
occurring in the proof are definable in $\mathbf{R}$. These sets are known
to have a simple structure, and there would be no difficulty in formalizing
the proof in a very small part of Peano arithmetic.

As a consequence of theorem 1 we obtain:

\begin{corollary}
For any direction $x_{0}$ there is a finite, fixed size set $\Gamma \subset 
\mathbb{S}_{+}^{2}$ which include $x_{0}$, the $e_{i}$'s and $b_{ij}$'s such
that any $p\in \mathcal{F}_{0}(\mathcal{\Gamma })$ satisfy $\left|
p(x_{0})\right| \leq \frac{1}{2}\left| p(x)\right| $ for some $x\in \Gamma $.
\end{corollary}

Here $\mathcal{F}_{0}(\mathcal{\Gamma })$ is the set of functions $p:\Gamma
\rightarrow \mathbb{R}$ which satisfy the conditions $p(e_{i})=p(b_{ij})=0$
and $p(x)+p(y)+p(z)=0$ for every orthogonal triple $x,y,z\in \Gamma $.
Iteration of this process yields a constructive proof of Gleason's theorem.
It would be nice to give an explicit construction of the set $\Gamma $ with
this property. As noted, there is a general algorithm to find it, but such
algorithm may not be feasible to execute in practice since the decision
procedure for $\mathbf{R}$ has a worst case {doubly} exponential time lower
bound, see Basu (1999).

We shall end with a few comments on the boundedness condition in Gleason's
theorem. Suppose that we drop the requirement that the functions $p\in 
\mathcal{F}_{0}$ are bounded, then the conclusion $p=0$ is simply false. One
can see this from the following cardinality consideration:

Denote by $\mathcal{G}$ the class of real functions $f$ satisfying $%
f(x)+f(y)+f(z)=0$ for every orthogonal triple $x,y,z\in \mathbb{S}_{+}^{2}$.
By Gleason's theorem the cardinality of all \emph{bounded} functions in $%
\mathcal{G}$ is $2^{\aleph _{0}}$. Note that $\mathcal{G}$ (including its
unbounded functions) is closed under composition with any additive group
homomorphism $\mathbb{R}\rightarrow \mathbb{R}$; i.e. if $g\in \mathcal{G}$
and $h:\mathbb{R}\rightarrow \mathbb{R}$, $h(x+y)=h(x)+h(y)$, then $h\circ
g\in \mathcal{G}$. Let $p$ be a bounded element of $\mathcal{G}$ given by
some non-zero matrix with trace zero. If one assumes the axiom of choice,
there are $\beth _{2}$ additive group homomorphisms from $\mathbb{R}$ to $%
\mathbb{R}$ . Composing them with $p$, we see that $\mathcal{G}$ is
enormous, and so certainly Gleason's theorem fails without the boundedness
requirement.

On the other hand consider a model of set theory in which the axiom of
choice fails, Solovay's (1970) model, or if every set has the property of
Baire say. Then every function $\mathbb{R}\rightarrow \mathbb{R}$ is
continuous on a set whose complement is meager; if it is also a group
homomorphism, it will be continuous everywhere. Thus, with this kind of
negation of the axiom of choice, it is not implausible that the boundedness
of frame functions might be automatic. Coming back to the remark at the end
of section 1, we also note that in such models there are no large $\Omega
\subset \mathbb{S}^{2}$ on which 0-1 valued states exist (Shipman 1990).

It seems instructive at all events to compare Lemma 3 to what one obtains
when $Y$ satisfies the stronger assumption, that \emph{every }function
(bounded or otherwise) vanishes. Then one can replace in Lemma 3(2) the
reference to $S_{k}(n)$ by

\begin{equation*}
\mathcal{S}_{\infty }(n)=\{S\subset \{0,\ldots ,n\}^{m}:\mathbf{R}_{-}\vdash
(\forall \alpha _{0})\ldots (\forall \alpha _{n})\left( \bigwedge_{s\in
S}(\sum_{i=1}^{m}\alpha _{s(i)}=0)\rightarrow (\alpha _{0}=0)\right) \}
\end{equation*}
The proof is the same as that of Lemma 3, mutatis mutandis, noting that when 
$B$ is a divisible Abelian group and $0\neq b\in B$, there exists a
homomorphism $H:B\rightarrow \mathbf{R}$ with $H(b)\neq 0$ (not necessarily
order preserving.)

\begin{acknowledgement}
IP acknowledges the support of the Israel Science Foundation (grant 879/02).
\end{acknowledgement}

\begin{center}
\bigskip

{\Large References}
\end{center}

Appleby, D.M. (2002) ``Existential Contextuality and the Models of Meyer,
Kent and Clifton'' \emph{Phys. Rev. A 65}, 022105.

Basu, S. (1999) New results on quantifier elimination over real closed
fields and applications to constraint databases \emph{Journal of the ACM 46}%
, 537--555.

Bell ,J. L. (1996) ``Logical reflections on the Kochen-Specker Theorem'' in
Clifton, R. (ed) \emph{Perspective on Quantum Reality} pp 227-235,
Dordrecht, Kluwer.

Bell, J. S. (1966): ''On the Problem of Hidden Variables in Quantum
Mechanics'', \emph{Reviews of Modern Physics 38}, 447-452; reprinted in
Bell, J S. (1987): \emph{Speakable and Unspeakable in Quantum Mechanics}
(Cambridge; Cambridge University Press)

Billinge, H. (1997) ``A Constructive Formulation of Gleason's Theorem'' 
\emph{Journal of Philosophical Logic 26}, 661-670.

Breuer, T. (2002) ``Another No-Go Theorem for Hidden Variable Models of
Inaccurate Spin Measurement'' \emph{Philosophy of Science Association 18th
Biennial Meeting - PSA \ \ 2002}, http://philsci-archive.pitt.edu/view
-200204.html.

Clifton, R. (1993) ``Getting Contextual and Nonlocal Elements-of-Reality the
Easy Way'' \emph{American journal of Physics 61}, 443-447.

Clifton, R. and Kent, A. (2000) ``Simulating Quantum Mechanics by
Non-Contextual Hidden Variables'' \emph{Proceedings of the.Royal Society of
London A 456}, 2101-2114.

Cooke, R. Keane, M. and Moran, W. (1984) ``An Elementay Proof of Gleason's
Theorem'' \emph{Mathematical Proceedings of the Cambridge Philosophical
Sosiety 98}, 117-128.

Fine, A. and Teller, P. (1978) Algebraic Constraints on Hidden Variables. 
\emph{Foundations of Physics 8} 629-636.

Gleason, A. M. (1957) Measures on the Closed Subspaces of a Hilbert Space. 
\emph{Journal of Mathematics and Mechanics 6}\textit{, }885-893.

Held, C. (2000) ``The Kochen and Specker Theorem'' \emph{Stanford
Encyclopedia of Philosophy} http://plato.stanford.edu/

Hellman, G. (1993) ``Gleason's Theorem is Not Constructively Provable'' 
\emph{Journal of Philosophical Logic 22}, 193-203.

Kochen, S. and Specker, E. P. (1967) ``The Problem of Hidden Variables in
Quantum Mechanics''. \emph{Journal of Mathematics and Mechanics 17}, 59-87.

Meyer, D. A. (1999) ``Finite Precision Measurement Nullifies the
Kochen-Specker Theorem'' \emph{Physical.Revew.Letters 83}, 3751-3754\emph{.}

Piron, C. (1976) \emph{Foundations of Quantum Physics}, (Reading MA,
Addison-Weseley).

Pitowsky, I. (1982) ''Substitution and Truth in Quantum Logic'', \emph{%
Philosophy of Science 49}, 380-401.

Pitowsky, I. (1983) ``Deterministic Model of Spin and Statistics'', \emph{%
Physical Review D27}, 2316-2326.

Pitowsky, I. (1985) ``Quantum Mechanics and Value Definiteness'', \emph{%
Philosophy of Science 52}, 154-156.

Pitowsky, I. (1998).``Infinite and Finite Gleason's Theorems and the Logic
of Indeterminacy'', \emph{Journal of Mathematical Physics 39}, 218 - 228.

Pitowsky, I. (2003) ``Betting on the Outcomes of Measurements: A Bayesian
Theory of Quantum Probability'' \emph{Studies in the History and Philosophy
of Modern Physics} \emph{34}, 395-414.

Richman, F. and Bridges, D. (1999) ``A Constructive Proof of Gleason's
Theorem \emph{Journal of Functional Analysis 162}, 287-312.

Shipman, J. (1990) ``Cardinal Conditions for Strong Fubini Theorems'' \emph{%
Transactions of the American Mathematical Society 321}, 465-481.

Shoenfield, J. R. (1967) \emph{Mathematical Logic }(Reading MA,
Addison-Weseley).

Solovay, R. M. (1970) ``A model of set-theory in which every.set of reals is
Lebesgue measurable'', \emph{Annals of Mathematics 92}, 1-56.

\end{document}